\documentclass[12pt]{amsart}
\usepackage[margin=1in]{geometry}                
\usepackage{graphicx}
\usepackage{amssymb,amsmath}
\usepackage{psfrag}
\newcommand{\rf}[1]{(\ref{#1})}
\newcommand{\sech}{\mathop{\rm sech}\nolimits}

\newcommand{\difd}{\mathrm{d}}
\newcommand{\expe}{\mathrm{e}}

\newcommand{\lcor}{L_{\mathrm{c}}}

\newcommand{\Xdepth}{X_{\mathrm{dep}}}
\newcommand{\Xtra}{X_{\mathrm{tra}}}
\newcommand{\Xint}{X_{\mathrm{int}}}
\newcommand{\Xstay}{X_{\mathrm{fluc}}}
\newcommand{\Xlinfirst}{X_{\mathrm{lin}}}

\newcommand{\pXtra}{p_{\Xtra}}
\newcommand{\pXint}{p_{\Xint}}

\begin{document}

\title{Random Polarization Dynamics in a Resonant Optical Medium}

\author[K.~Newhall, E.~Atkins, P.~Kramer, G.~Kova\v{c}i\v{c}, I.~Gabitov]{Katherine A. Newhall,$^1$ Ethan P. Atkins,$^2$  Peter R. Kramer,$^3$ Gregor Kova\v{c}i\v{c},$^{3,*}$ \\ and Ildar R. Gabitov$^{4}$\\
{}\\
$^1$Courant Institute of Mathematical Sciences, New York University\\
$^2$ Department of Mathematics, University of California, Berkeley\\
$^3$Mathematical Sciences Department, Rensselaer Polytechnic Institute\\
$^4$Department of Mathematics,  University of Arizona\\
$^*$Corresponding author: kovacg@rpi.edu
}

\date{\today}

\begin{abstract}
Random optical-pulse polarization switching along an active optical medium in the $\Lambda$-configuration with spatially disordered occupation numbers of its lower energy sub-level pair is described using the idealized integrable Maxwell-Bloch model.   Analytical results describing the light polarization-switching statistics for the single self-induced transparency pulse are compared with statistics obtained from direct Monte-Carlo numerical simulations.
\end{abstract}

\maketitle


The model of light interacting with a material sample composed of three-level active atoms has made possible the descriptions of several nontrivial optical phenomena, including lasing without inversion~\cite{PhysRevLett.62.1033}, slow light~\cite{hau99}, and electric-field polarization of solitons in self-induced transparency~\cite{Basharov90}.    Its simplest version including a non-degenerate upper and two degenerate lower working atomic levels --- the $\Lambda$ configuration  --- is completely integrable when the pulse width is much shorter than the medium relaxation times~\cite{maimistov84}.  It describes a new type of a self-induced transparency pulse, which may be a solitary wave only asymptotically, but in general switches into one of the two purely two-level transitions between one of the lower levels and the upper level. These transitions correspond to circularly polarized light, and the direction of the switching is determined by the population sizes of the degenerate lower levels~\cite{Maimistov85,Byrne03}.   Thus, for spatially disordered populations, random polarization switching takes place as a light soliton travels along the material sample.   The integrability of the  $\Lambda$ configuration furnishes a unique opportunity to study the mechanism responsible for this random switching and its statistical properties   exactly in the framework of a sufficiently idealized model, which otherwise would be impossible because of strong nonlinearity.   In this letter, we both discuss the analytical results~\cite{akkg12} on this switching and compare them with the results of numerical simulations.

Resonant propagation of ultra-short, monochromatic, elliptically polarized light pulses through a two-level, active medium with a doubly degenerate ground level ($\Lambda$-configuration)
is described by the quasi-classical  Maxwell-Bloch system~\cite{konopnicki81,maimistov84,Basharov90,Byrne03}
\begin{subequations}
\begin{align}
\partial_t E_\pm+\partial_x E_\pm & = 
\int_{-\infty}^{\infty} \rho_\pm
\, g(\nu) \difd \nu , \label{erhoeqn} \\
\partial_t \rho_+-2i \lambda \rho_+ & = 
\left[E_+({\mathcal N}-n_+) -
E_-\mu^*\right]/2,\label{rhoPeqn} \\
\partial_t \rho_- -2 i \lambda \rho_- & =  
\left[E_-({\mathcal N}-n_-)
- E_+ \mu \right]/2,\label{rhoMeqn} \end{align}
\begin{align}
\partial_t \mu & =  \left[{E_+}^* \rho_- + E_-
{\rho_+}^* \right]/2,\\
\partial_t {\mathcal N} & = - \left[E_+{\rho_+}^* +
{E_+}^*\rho_+  + E_- {\rho_-}^* + {E_-}^*
\rho_- \right]/2, \\
\partial_t n_{\pm} & = \left[E_{\pm}
{\rho_\pm}^* + {E_{\pm}}^* \rho_\pm\right]/2.
\end{align}
\label{lambdaeqns}
\end{subequations}
Here, $E_\pm(x,t)$ are the envelopes of the electric field and  $\rho_\pm(x,t,\lambda)$ and $\mu(x,t,
\lambda)$ of the 
medium-polarization, $n_{\pm}(x,t,
\lambda)$ and ${\mathcal N}(x,t, \lambda)$ the
population densities of the ground
and excited levels, respectively, $\lambda$ the frequency detuning, and
$g(\lambda)\geq 0$, with $\int_{-\infty}^\infty g(\lambda) \,\difd\lambda =
1$, the spectral-line shape.    The ``$+$'' and ``$-$'' transitions interact with the left- and right-circularly polarized pulse components, while $\mu$ is due to the
two-photon transition between the ground levels.  The purely two-level ``$+$'' and ``$-$'' transitions are invariant and involve only circularly-polarized light.  A time-conserved quantity of Eqs.~(\ref{lambdaeqns}) is ${\mathcal N}+n_++n_-=1$,  where unit normalization is chosen.

The approximations made in Eqs.~(\ref{lambdaeqns}) are (i) the pulse-width is much longer than the light oscillation period (slowly-varying envelope approximation) and much shorter than the relaxation time-scales in the medium, and (ii) unidirectional propagation.  The latter holds provided the interaction time of counter-propagating pulses is much shorter than the nonlinear-response time of the medium.   Equations~(\ref{lambdaeqns}) are dimensionless, e.g., the speed of
light is $ c=1 $.

If the spectral width of the pump pulse priming the ground states of the medium is much broader than the width of $g(\lambda)$,  i.e., the initial populations can be considered homogeneous within
 the width of $g(\lambda)$, we find the two components of the soliton solution~\cite{Maimistov85,Byrne03,akkg12}
\begin{align}\label{onesoliton}
&E_\pm(x,t)=4i\beta G_\pm(x) e^{i\Theta_\pm(x,t)}
\sech\biggl[2 \beta (t-x) + \tau  x \nonumber\\ &
+ \frac{1}{2} \ln \frac{|d_+||d_-|}{2 \beta^2}
+ \frac{1}{2} \ln \cosh \left(2 \tau  A(x) + \ln \frac{\left\vert
d+\right\vert}{\left\vert d_-\right\vert}\right)\biggr],
\end{align}
where 
$$
G_\pm(x)=\sqrt{\left[1\pm\tanh\left(2\tau A(x)+\ln\left\vert
d_+\right\vert/\left\vert d_-\right\vert\right)\right]/2}
$$
are their amplitudes and  
$$
\Theta_\pm (x,t) = 2 \gamma (t-x) + \sigma [x\pm A(x)]
-\arg d_\pm
$$
their phases, which in turn depend on the soliton parameters $\gamma$ and $\beta$.
Here, 
\begin{equation}
\label{Adef}
A(x)=\textstyle \int_0^x\alpha(\xi)\difd\xi,
\end{equation}
is the cumulative initial population difference $\alpha(x)$ along the medium sample up to any given position $x$, which satisfies the asymptotic condition
\begin{equation}
\label{alphadef} \lim_{t \rightarrow
-\infty} n_\pm (x,t,\lambda)=\left[ 1\pm\alpha(x)\right]/2\geq 0.\end{equation}
The rest of the material variables are known to vanish as $t\to-\infty$ for this solution~\cite{Byrne03}, so that only the two degenerate lower levels are populated initially. 
Putting the initial time at $-\infty$ is  justified because, in gases, the lifetime of the system ranges from $10^{-5}$ to $10^{-3}$ seconds, while the typical pulse-width is $10^{-8}$ seconds or shorter~\cite{allen87}.
The real-valued coefficients $\sigma$ and $\tau$ are given by
\begin{equation}
\label{Gdef}
 \sigma+i\tau=
 \int_{-\infty}^{\infty} \frac{g(\nu)}{8(\gamma+i\beta-\nu)} \difd \nu,
\end{equation}
with $\beta>0$.  
Equation~(\ref{onesoliton}) shows that the maximal amplitude of each soliton component equals $4\beta$ and its temporal width 
equals $1/(2\beta)$.  The constants $d_\pm$  give the soliton phase and position.  Note that, since $\tau<0$, the amplitude $G_+(x)$ decreases and $G_-(x)$ increases with increasing $A(x)$, and vice versa with decreasing $A(x)$, which is the polarization-switching effect of~\cite{Maimistov85,Byrne03}.

The light-pulse polarization can be described in terms of the polarization ellipse, which is characterized by the orientation and ellipticity angles, $\psi$ and  $\eta$, with $-\pi/4\leq \eta\leq \pi/4$.   These can be found from the formulas  
$$
\tan 2\psi = i(E_+E_-^*-E_-E_+^*)/(E_+E_-^*+E_-E_+^*)
$$ 
and 
$$\sin2\eta=(|E_+|^2 - |E_-|^2)/(|E_+|^2 + |E_-|^2),$$
 which, for the soliton (\ref{onesoliton}) give~\cite{Byrne03}
$$ \psi = -\sigma A(x) +\arg \left(d_-^* d_+\right)/2,$$  
$$\sin 2\eta =
\tanh \left[ 2\tau A(x) +
\ln\left(\left\vert d_+\right\vert/\left\vert d_-\right\vert\right)
\right].$$ 
Note that these two angles are time-independent.

If the initial population difference $\alpha(x)$ in the medium is random and spatially statistically homogeneous, we can approximate it as white noise
\begin{equation}\label{eq:whitemodel}
\langle \alpha (x) \rangle  = b, \quad
\langle [\alpha (x) - b] [\alpha (x')-b]
 \rangle = a^2\delta(x-x'),
 \end{equation}
where $\langle\cdot\rangle$ denotes ensemble averaging over all possible realizations of $\alpha(x)$, and $\delta(\cdot)$ is the Dirac Delta function.   This approximation is consistent provided the pulse-carrier frequency $\lambda_0$, the correlation length $L_c$ of $\alpha(x)$, the soliton width $1/\beta$, and the observation location $x$ along the sample satisfy the inequalities
$\lambda_0\ll \lcor\ll 1/\beta\ll x $.   The first is related to the slowly-varying envelope approximation (mentioned above), the second to the unidirectionality assumption, and the last to the white-noise assumption.  In this case, the cumulative integral $A(x)$ in Eq.~\rf{Adef} can be approximated as $
A(x)\sim aW(x)+bx$, where $W(x)$ is the standard Wiener process. Note that the parameter $a$ in Eq.~\rf{eq:whitemodel}, the correlation length $\lcor$, and the variance $\sigma^2_\alpha$ of a true  initial population difference $\alpha(x)$ are related by
$ a = \sqrt{2\lcor}\, \sigma_\alpha$.

In an experiment,  $L_c$  would be approximately the same as the coherence length 
$
\ell_c \sim \lambda_p^2/\Delta \lambda_p
$
 of the pump light used to prepare the optical medium,   
where $\lambda_p$ is the average wavelength of the pump light and $\Delta \lambda_p$ the characteristic width of the light-source spectral line.  If the pump was a Ti-sapphire laser,  $\lambda_p \sim 800$ nm and $\Delta \lambda_p \sim 5$ nm~\cite{wolf07}, so $\ell_c \sim 0.1 \text{ mm}\gg \lambda_0 $ ($\sim 600$ nm for sodium vapor)   and a
several-centimeters long experimental device would be  sufficiently long  to capture the desired statistical effects. 

\begin{figure}
\includegraphics[scale=0.4]{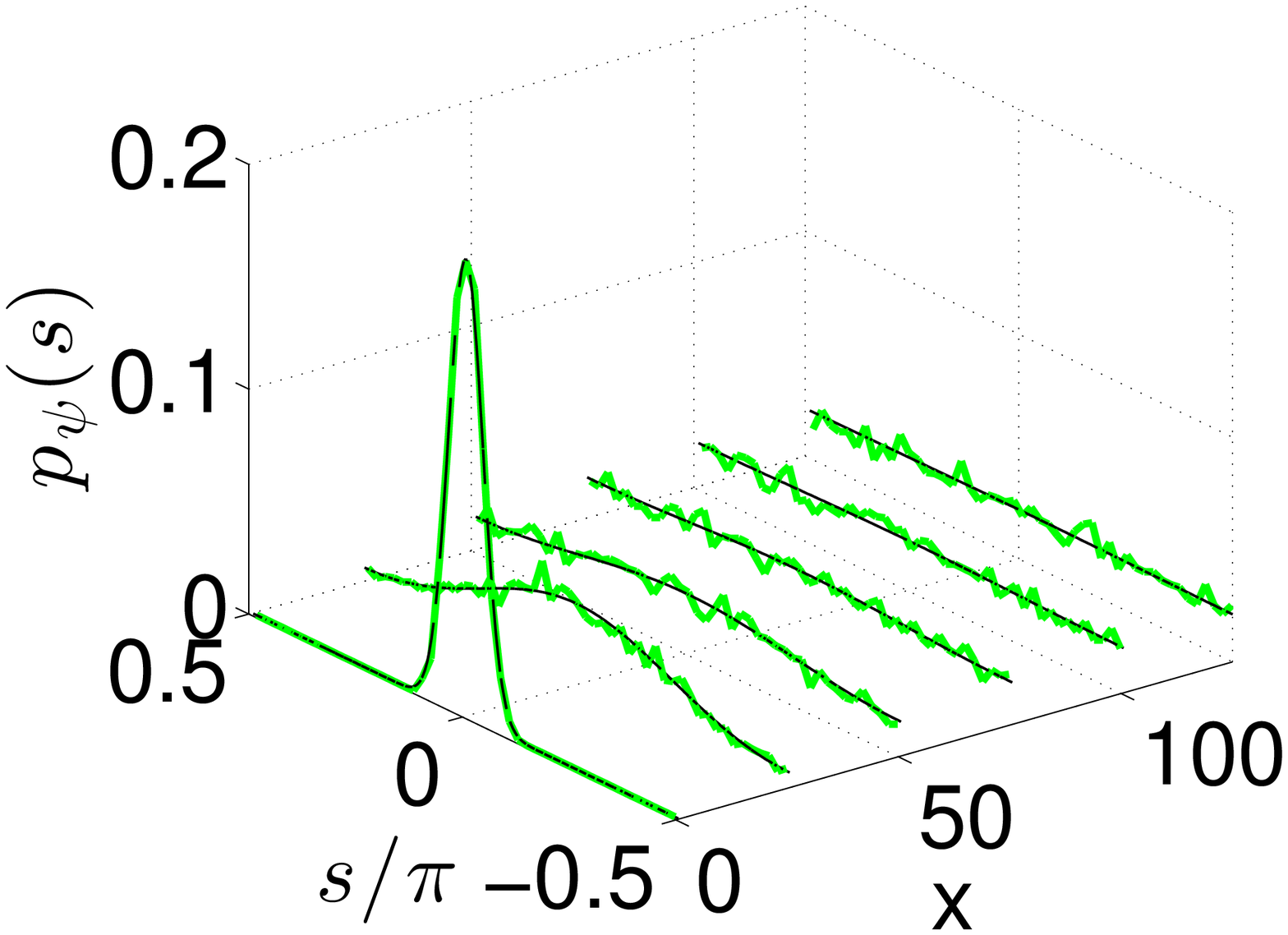}
\includegraphics[scale=0.4]{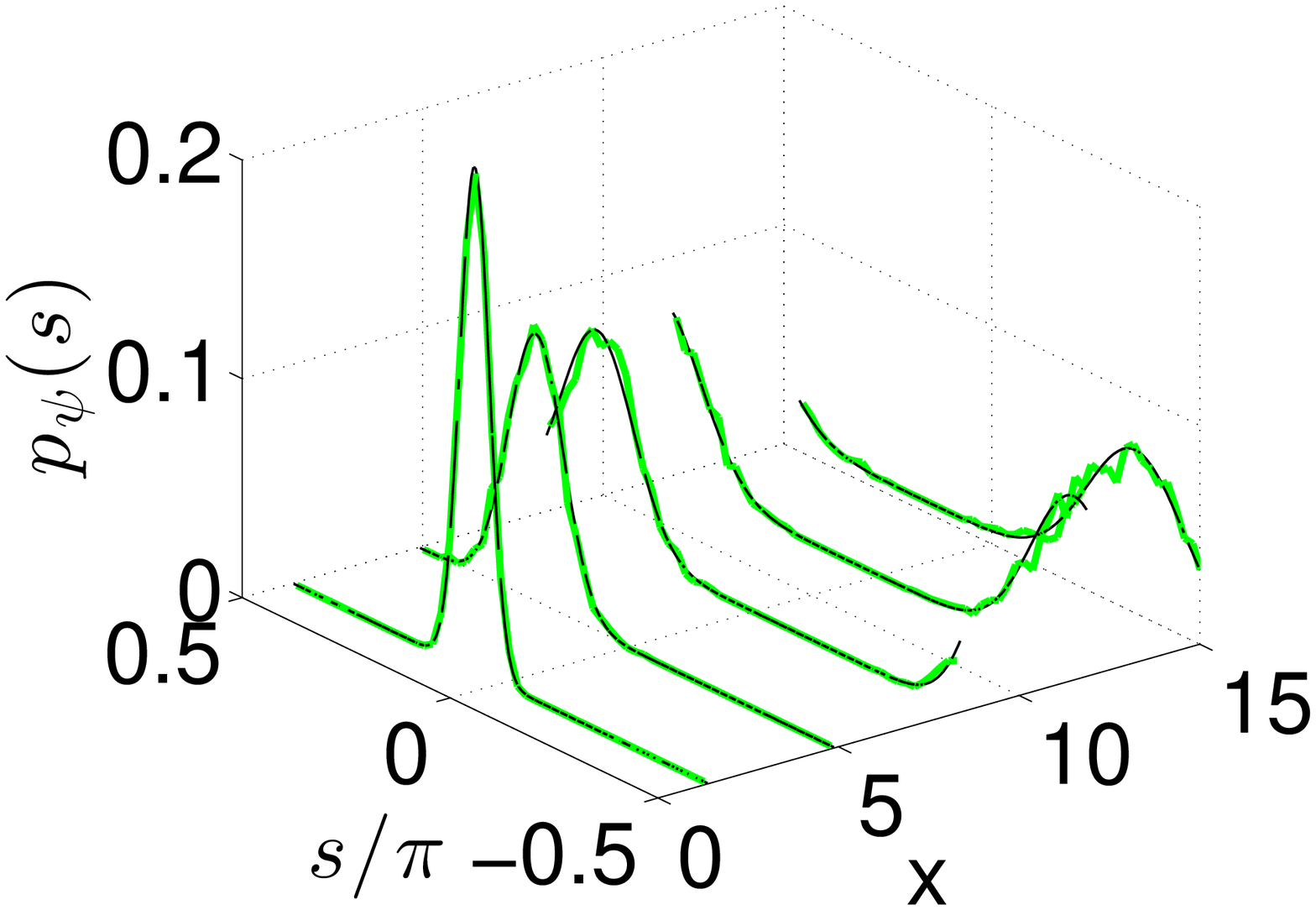}
\caption{PDF $p_\psi(x;s)$, with $\beta=1/3$, $\gamma=1/3$, $\varepsilon=0$, $d_+=d_-=i$, theoretical (black lines) and results from 1600 simulations (gray lines; green online).  Left: $b=0$, $a=0.75$.  Right: $b=-0.75$, $a=0.5$.}
\label{fig:ppsi}
\includegraphics[scale=0.4]{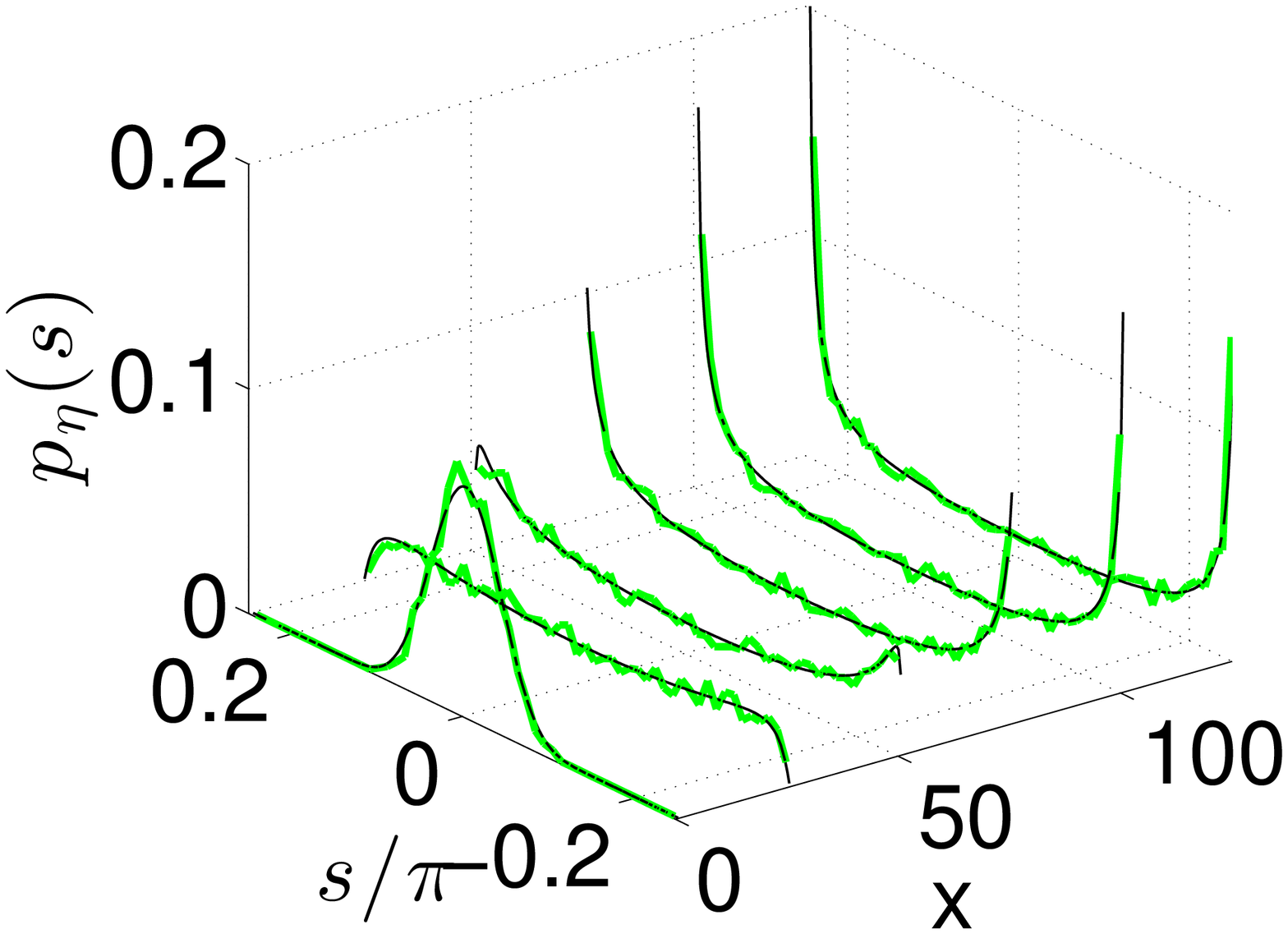}
\includegraphics[scale=0.4]{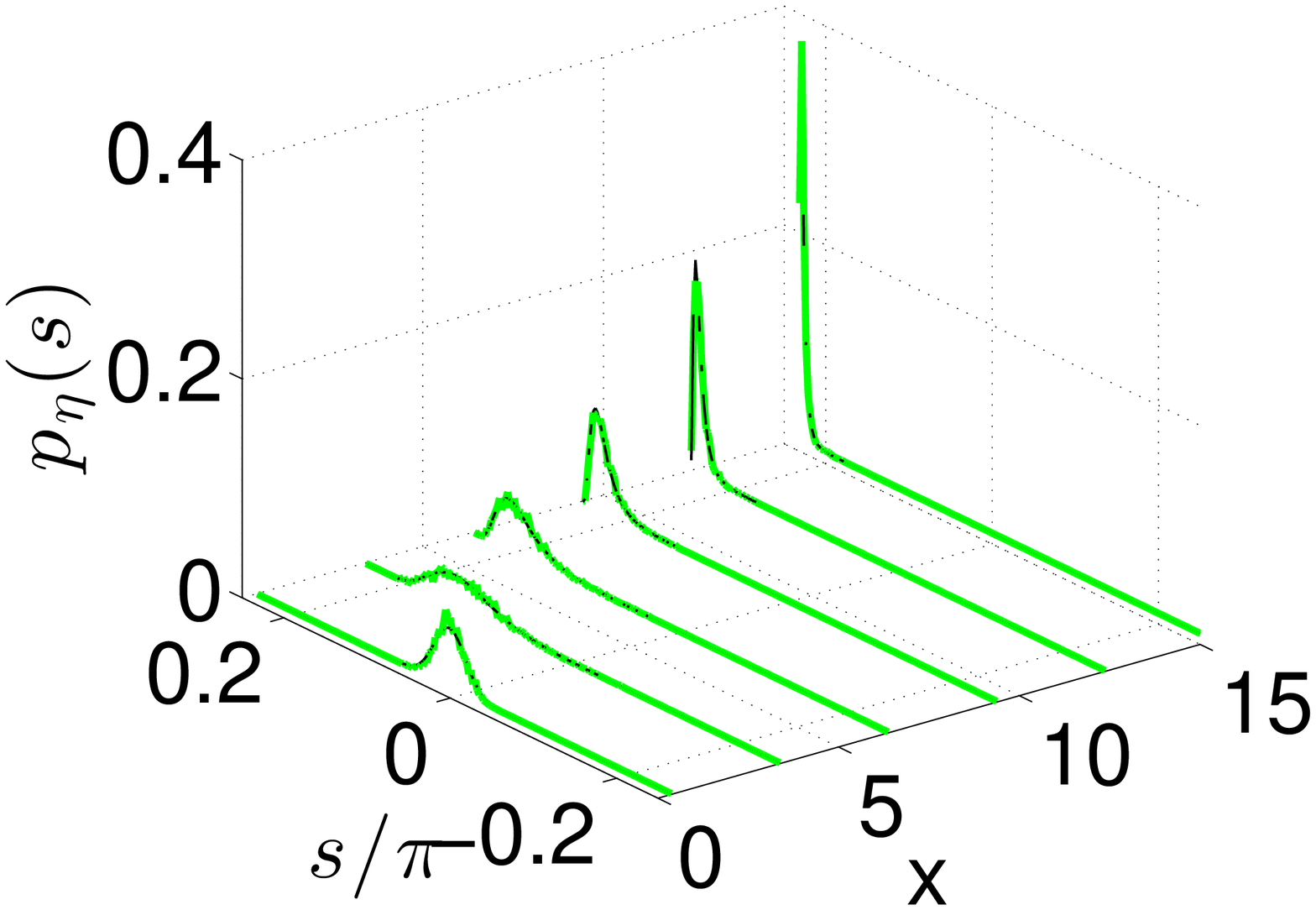}
\caption{PDF $p_\eta(x;s)$, with $\beta=1/3$, $\gamma=1/3$, $\varepsilon=0$, $d_+=d_-=i$, theoretical (black lines) and results from 1600 simulations (gray lines; green online).  Left: $b=0$, $a=0.75$.  Right: $b=-0.75$, $a=0.5$.}
\label{fig:peta}
\end{figure}

The orientation angle $ \psi (x) $ behaves like
a Brownian motion with drift $ - \sigma  b $ and diffusion coefficient $ \frac{1}{2} \sigma ^2 a^2 $and its probability density function (PDF), $p_{\psi}(x;s)$, at any $ x $ is Gaussian in $s$ 
with mean
$
\langle \psi (x)\rangle= -\sigma bx +\frac{1}{2}\arg(d^*_-d_+)
$ 
and variance
$
\sigma^2_\psi (x) =\sigma^{2}a^{2}x$.  
Note that the value of $\psi(x)$ is fixed at $\psi=\arg(d^*_-d_+)/2$ when $\sigma=0$.  The PDF $p_{\psi}(x;s)$ shows excellent agreement with numerical simulations in Fig.~\ref{fig:ppsi}.   The Lorentzian spectral-line shape $g(\lambda)=\varepsilon/\pi(\lambda^2+\varepsilon^2)$ was used in determining $p_{\psi}(x;s)$ and all subsequent PDFs.  Comparisons in all figures are made with $\varepsilon=0$ corresponding to the Dirac Delta function, i.e., the limit of an infinitely sharp spectral line.

\newpage

The PDF for the ellipticity angle $\eta$ at any $ x $ equals
$$ p_{\eta}(x;s)=(1/\sqrt{2\pi x}\, a|\tau|\cos 2s) 
 \exp\{-[\tanh^{-1}(\sin 2s)-2\tau bx-\ln|d_+|/|d_-|]^{2}/8a^{2}\tau^{2}x\}$$ for $-\pi/4\leq s\leq \pi/4$.
For large $x$,  $p_\eta(x;s)$  concentrates at one or both circular polarizations at $ s = \pm \pi/4 $ as $p_{\eta}(x;s)\sim[\delta(s+\pi/4)+\delta(s-\pi/4)]/2$  if  $b=0$, $\sim \delta(s+\pi/4)$ if $b>0$ and $\sim \delta(s-\pi/4)$ if $b<0$. The PDF $p_{\eta}(x;s)$ shown in Fig.~\ref{fig:peta} accurately describes the numerical simulations.

To quantify the polarization switching statistics, we consider the periods during which the ellipticity angle $\eta$ enters, exits, and stays in the vicinity of either circular polarization, $\pi/4-\kappa \equiv \eta_\kappa<|\eta|<\pi/4$, for some appropriately chosen small angular distance $\kappa$.  More precisely, we study the statistics of the following random distances: $\Xtra $, over which the pulse polarization evolves from linear ($ \eta = 0 $) to nearly-circular of either orientation  ($|\eta| = \eta_\kappa $); $\Xint$,  over which the pulse polarization evolves from linear to nearly-circular of either orientation and back to linear; and
$\Xdepth$, beyond which the pulse polarization remains forever circularly-polarized,  $ |\eta| > \eta_\kappa $, for all greater distances.

\begin{figure}[t]\begin{center}
\includegraphics[scale=0.4]{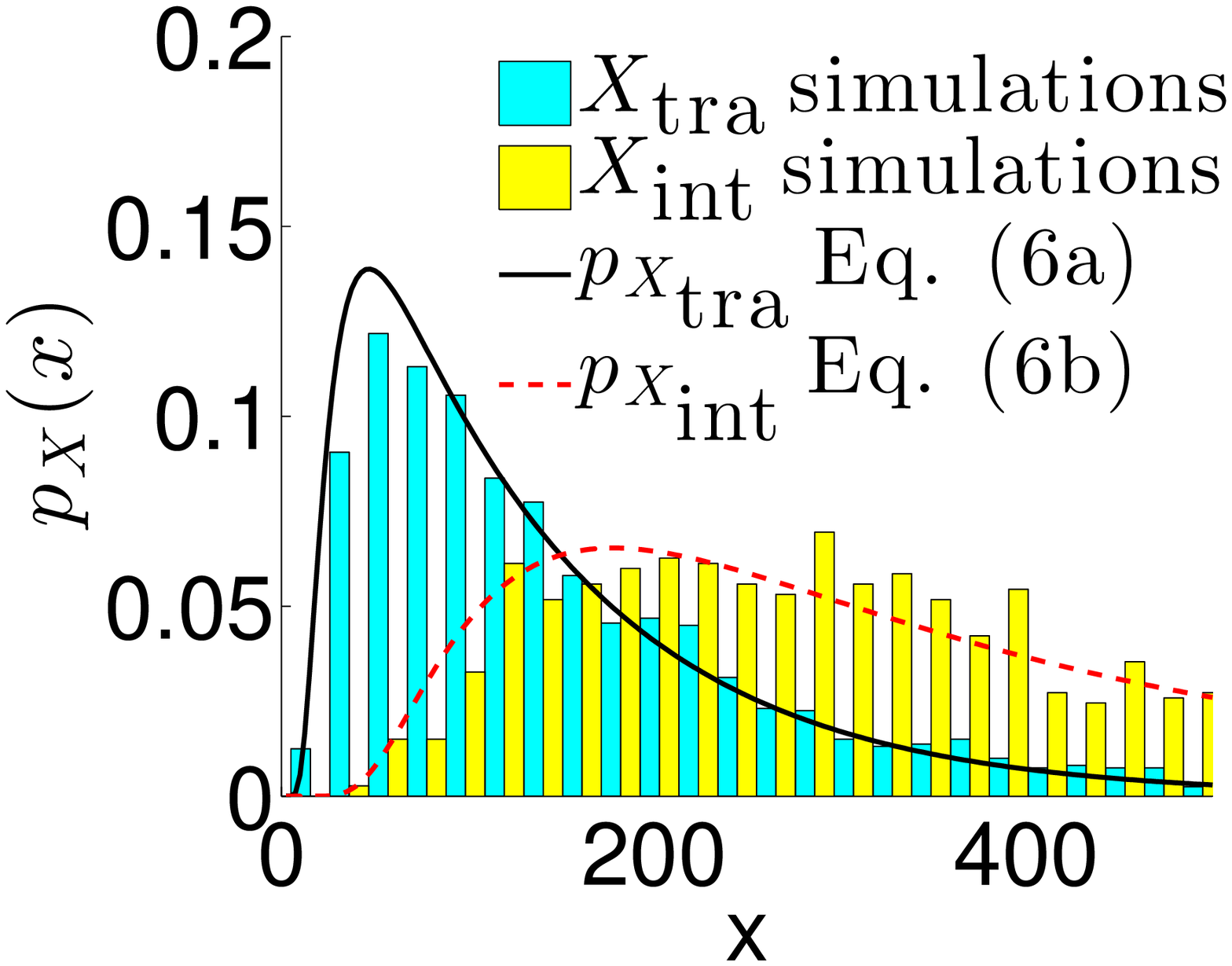}\includegraphics[scale=0.4]{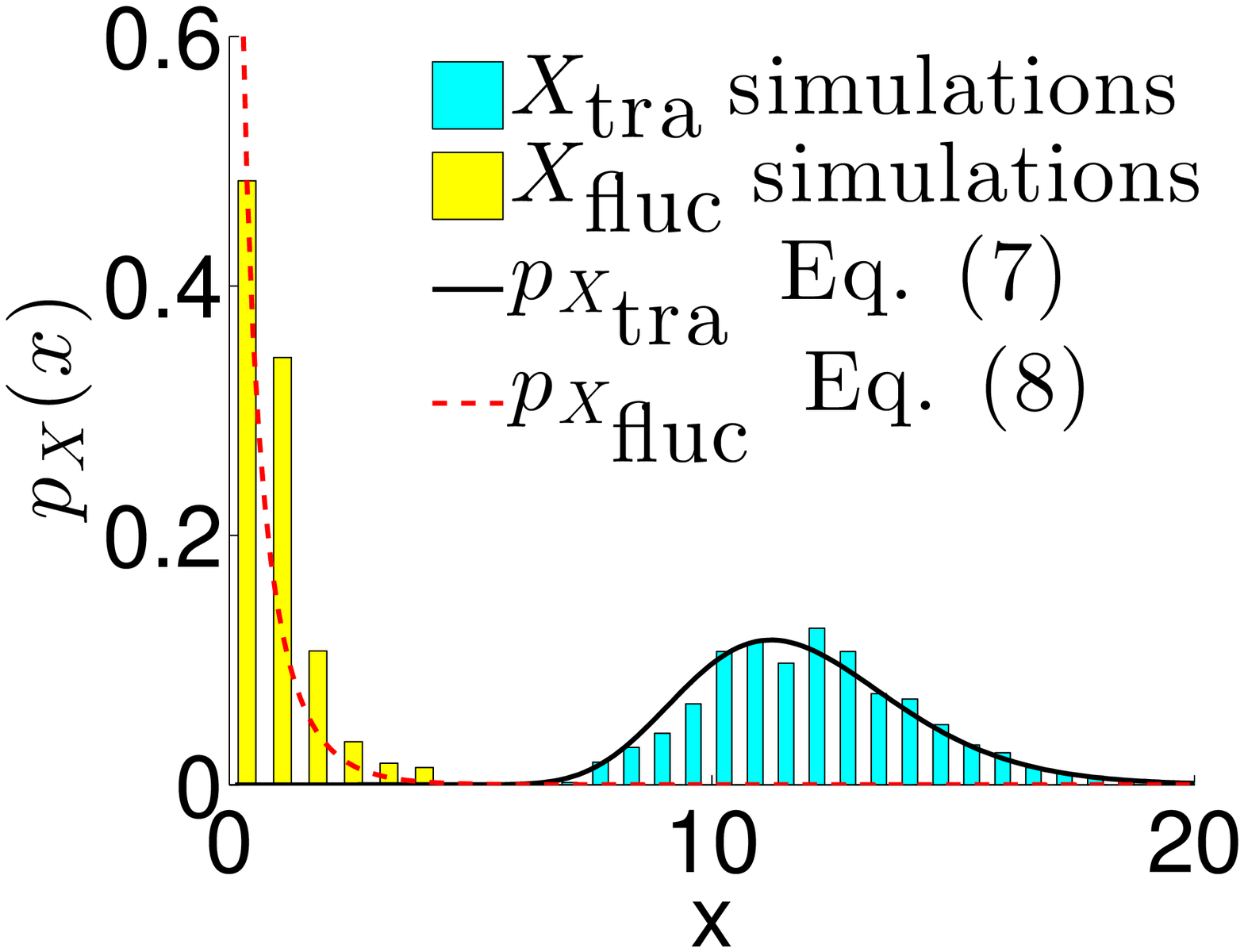}
\caption{Comparison of exit time statistics between analytical formulas and results from numerical simulations when (left) no bias, $b=0$, $a=0.75$, and (right) $b=-0.75$, $a=0.5$.  For both, $\beta=1/3$, $\gamma=1/3$, $\varepsilon=0$, $d_+=d_-=i$.}
\label{fig:pxs}
\end{center}
\end{figure}

When $ b= \langle \alpha (x) \rangle = 0 $
we find~\cite{akkg12}
\begin{subequations} \label{eq:pxs}
\begin{align}
\pXtra (x) &= \sqrt{2L/\pi x^3} 
 \sum_{n=-\infty}^{\infty}
(4n+ 1) \expe^{- (4n+1)^2L/2x},\\
\pXint (x) &= \sqrt{2L/\pi x^3} 
\left[ \sum_{n=0}^{\infty}
(4n+ 2) \expe^{- (4n+2)^2L/2x} -  \sum_{n=1}^{\infty}
4n  \expe^{- 16 n^2 L/2x}\right] , \label{eq:pxint}
\end{align}
\end{subequations}
where
$
L= \left[\tanh^{-1}( \cos 2 \kappa)/2a |\tau|\right]^2$.  These two distributions are depicted in Fig.~\ref{fig:pxs}(left) and agree with the numerical simulations.  Note that $\langle \Xtra \rangle = L$,  
$\sigma^2_{\Xtra} = 2 L
^2/3$, but $\langle\Xint\rangle=\infty$.

Another statistic, when $b=0$, is the fraction $\Phi$ of the length over which the polarization $ \eta $ takes a certain sign.  Its PDF is $p_{\Phi} (\phi) = [\pi \sqrt{\phi(1-\phi)}]^{-1}$~\cite{akkg12}.

For $ b = \langle \alpha (x) \rangle \neq 0$, if $ b $ has opposite sign to the  ellipticity angle 
$$\eta_0 = \sin^{-1} \left[\tanh \left(\ln (|d_+|/|d_-|)\right)\right]/2$$ of the injected pulse,  
the soliton polarization will first become linear after a distance $ \Xlinfirst $, move near the favored circular polarization over a subsequent distance $ \Xtra $,
and
never leave this ultimate circular polarization after a subsequent distance $ \Xstay $. Thus, $ \Xdepth = \Xlinfirst + \Xtra + \Xstay $.  
The first two distances are distributed as~\cite{akkg12}
\begin{equation}
p_X (x) = \left(|b| \ell/a \sqrt{2 \pi x^3}\right) \exp\left[- b^2 (\ell - x)^2/2 a^2 x\right],
\end{equation}
and $ \langle X \rangle = \ell$,  
$\sigma^2_X =  a^2 \ell /b^2$, with 
$$
\ell = \ln (|d_+|/|d_-|) /2 |\tau  b| 
$$
for $ X= \Xlinfirst $ and 
$$ \ell=\tanh^{-1} (\cos 2 \kappa) /2 |\tau  b|$$
 for $X=\Xtra$. The probability density of $\Xstay $ is 
\begin{equation}
p_{\Xstay}(x) =(|b|/a\sqrt{2\pi x})\exp(-b^2 x/2a^2). \label{eq:xstay}
 \end{equation}
 When $ b $ and $ \eta_0 $ have the same sign, results are similar.  The above two distributions show excellent agreement with the numerical simulations in the right panel of Fig.~\ref{fig:pxs}.
  
 Numerical simulations of system~\rf{lambdaeqns} are carried out by aligning a grid in $x$ and $t$ with spacing $\Delta x$ and $\Delta t$ to the characteristics of Eq.~\rf{erhoeqn} (i.e. $\Delta x = \Delta t$) and implementing a scheme based on the implicit mid-point method.  The initial conditions in $n_{\pm}$ are obtained using Eq. (4) in which $\alpha(x)$ is a Gaussian random variable at each grid point with mean $b$ and standard deviation $a/\sqrt{\Delta x}$.  The soliton enters through the boundary condition on the left hand side, and no boundary condition is needed on the right hand side.
  
 In conclusion, we have shown excellent comparison between the analytical results and direct numerical simulations describing the statistics of the orientation, $\psi$, and eccentricity, $\eta$, of the light polarization ellipse.  We also found favorable comparison for the distance over which the light polarization evolves from linear to nearly-circular, $\Xtra$, and back again to linear, $\Xint$, and, for unequal distributions within the lower populations, the distance between the point at which light first comes close to its favored circular polarization and the point at which it finally forever remains near it,  $\Xstay$. 
 
We thank Misha Stepanov for useful discussions, and acknowledge partial support by NSF and DOE Graduate Fellowships, NSF grants DMS-0509589, DMS-0636358, and DMS-1009453, AROMURI award 50342-PH-MUR, and the Russian Federal Goal-Oriented Program ``SSEPIR.''

\bibliographystyle{plain}

\begin{thebibliography}{10}
\newcommand{\enquote}[1]{``#1''}

\bibitem{PhysRevLett.62.1033}
S.~E. Harris, Phys. Rev. Lett. \textbf{62}, 1033 (1989).

\bibitem{hau99}
L.~N. Hau, S.~E. Harris, Z.~Dutton, and C.~H. Behroozi, Nature \textbf{397},
  594 (1999).

\bibitem{Basharov90}
A.~M. Basharov, S.~O. Elyutin, A.~I. Maimistov, and Y.~M. Sklyarov, Phys Rep.
  \textbf{191}, 1 (1990).

\bibitem{maimistov84}
A.~I. Maimistov, Sov. J. Quantum Electron. \textbf{14}, 385 (1984).

\bibitem{Maimistov85}
A.~I. Maimistov and Y.~M. Sklyarov, Optics and Spectroscopy \textbf{59}, 459
  (1985).

\bibitem{Byrne03}
J.~A. Byrne, I.~R. Gabitov, and G.~Kova\v{c}i\v{c}, Physica D \textbf{186}, 69
  (2003).

\bibitem{akkg12}
E.~P. Atkins, P.~R. Kramer, G.~Kova\ifmmode \check{c}\else
  \v{c}\fi{}i\ifmmode~\check{c}\else \v{c}\fi{}, and I.~R. Gabitov, Phys. Rev.
  A \textbf{85}, 043834 (2012).

\bibitem{konopnicki81}
M.~J. Konopnicki and J.~H. Eberly, Phys. Rev. A \textbf{24}, 2567 (1981).

\bibitem{allen87}
L.~Allen and J.~H. Eberly, \emph{Optical Resonance and Two-Level Atoms} (Dover,
  New York, 1987).

\bibitem{wolf07}
E.~Wolf, \emph{Introduction to the Theory of Coherence and Polarization of
  Light} (Cambridge University Press, Cambridge, New York, 2007).

\bibitem{PhysRevLett.62.1033}
S.~E. Harris, \enquote{Lasers without inversion: Interference of
  lifetime-broadened resonances,} Phys. Rev. Lett. \textbf{62}, 1033--1036
  (1989).

\bibitem{hau99}
L.~N. Hau, S.~E. Harris, Z.~Dutton, and C.~H. Behroozi, \enquote{Light speed
  reduction to 17 metres per second in an ultracold atomic gas,} Nature
  \textbf{397}, 594--598 (1999).

\bibitem{Basharov90}
A.~M. Basharov, S.~O. Elyutin, A.~I. Maimistov, and Y.~M. Sklyarov,
  \enquote{Present state of self-induced transparency theory,} Phys Rep.
  \textbf{191}, 1--108 (1990).

\bibitem{maimistov84}
A.~I. Maimistov, \enquote{Rigorous theory of self-induced transparency in the
  case of a double resonance in a three-level medium,} Sov. J. Quantum
  Electron. \textbf{14}, 385--389 (1984).

\bibitem{Maimistov85}
A.~I. Maimistov and Y.~M. Sklyarov, \enquote{Coherent interaction of light
  pulses with a three-level medium,} Optics and Spectroscopy \textbf{59},
  459--461 (1985).

\bibitem{Byrne03}
J.~A. Byrne, I.~R. Gabitov, and G.~Kova\v{c}i\v{c}, \enquote{Polarization
  switching of light interacting with a degenerate two-level optical medium,}
  Physica D \textbf{186}, 69--92 (2003).

\bibitem{akkg12}
E.~P. Atkins, P.~R. Kramer, G.~Kova\ifmmode \check{c}\else
  \v{c}\fi{}i\ifmmode~\check{c}\else \v{c}\fi{}, and I.~R. Gabitov,
  \enquote{Stochastic pulse switching in a degenerate resonant optical medium,}
  Phys. Rev. A \textbf{85}, 043834 (2012).

\bibitem{konopnicki81}
M.~J. Konopnicki and J.~H. Eberly, \enquote{Simultaneous propagation of short
  different-wavelength optical pulses,} Phys. Rev. A \textbf{24}, 2567--2583
  (1981).

\bibitem{allen87}
L.~Allen and J.~H. Eberly, \emph{Optical Resonance and Two-Level Atoms} (Dover,
  New York, 1987).

\bibitem{wolf07}
E.~Wolf, \emph{Introduction to the Theory of Coherence and Polarization of
  Light} (Cambridge University Press, Cambridge, New York, 2007).

\end{thebibliography}

\end{document}